\documentclass[12pt]{article}

\usepackage{amsmath,amsfonts}
\usepackage{fancyhdr}
\usepackage{graphicx}
\usepackage{pstricks}

\font\titlefont=cmssbx10 at 18pt
\font\authorfont=cmr12
\font\abstractfont=cmr10
\font\abstractit=cmti10

\font\sectionfont=cmssbx10 at 14pt

\font\headerfont=cmss9
\font\headerit=cmssi9
\font\footerfont=cmss9

\begin{document}

\begin{center}
\begin{minipage}{3.5in}
\vskip45pt
\begin{center}
{\titlefont 
Difficulties with the Klein-Gordon Equation
}
\end{center}
\vskip12pt 
{\authorfont 
E. Comay \\
School of Physics and Astronomy \\
Raymond and Beverly Sackler Faculty of Exact Sciences \\
Tel Aviv University, Tel Aviv 69978, Israel \\
Email: elic@tauphy.tau.ac.il
}
\vskip18pt 
{\abstractfont
Relying on the variational principle, it is proved that new
contradictions emerge from an analysis of the Lagrangian density
of the Klein-Gordon field: normalization problems arise and interaction
with external electromagnetic fields cannot take place.
By contrast, the Dirac equation
is free of these problems. Other inconsistencies arise if the Klein-Gordon
field is regarded as a classical field.
}
\vskip12pt 
{\abstractfont
{\abstractit Keywords}: 
Relativistic Quantum Mechanics, Klein-Gordon Equation
}
\end{minipage}
\end{center}

\vskip12pt
{\noindent \sectionfont 1. Introduction}
\vskip6pt

The Klein-Gordon (KG) equation (called also Schroedinger's relativistic
wave equation)
\begin{equation}
(\Box + m^2)\phi = 0
\label{eq:KGEQ}
\end{equation}
was proposed by several authors in the very early days of quantum
mechanics (see [1], bottom of p. 25). Difficulties with
this equation were pointed out soon after its publication.
In particular, it was claimed that, as a second order equation
of the time derivative, solutions of the KG equation cannot represent
local probability of a particle (see [2] pp. 7,8).
These problems motivated Dirac
to construct his relativistic first order differential 
equation, now regarded as
the relativistic form of the Schroedinger equation for a spin-1/2
particle.

The debate surrounding the KG equation has gone on for many decades. Dirac
maintained that the KG equation is unacceptable[3] throughout his
life. A contradictory point 
of view argues that problems of the KG equation can be resolved and that
Dirac's point of view is incorrect (see [1], second column of p. 24). 
Today, the KG equation is
used as the equation of motion of a massive spinless particle
and it can be found in some books discussing classical field theory
([4]-[6])
and in many books on quantum field theory(see e.g. [2], p. 21, 
[6] and [7]).

This work uses units where $\hbar = c = 1$. In these units, a
shorthand notation of dimensions of physical quantities is used. 
As it turns out, this notation makes the discussion clearer. Thus,
mass, energy, momentum, electromagnetic potentials 
and acceleration have the dimension $[L^{-1}]$.
Length and time have the dimension $[L]$ and  
electric charge, velocity, angular momentum and action are
dimensionless. The square brackets used 
here avoid confusion between
the symbol of length and that of the Lagrangian.
The Lorentz metric
$g_{\mu \nu}$ is diagonal and its entries are 
(1,-1,-1,-1). The summation convention holds for a pair of upper 
and lower indices. The
lower case symbol $_{,\mu}$ denotes the partial differentiation with
respect to $x^\mu$.

The cornerstone of this work is the variational principle
which is used to prove new difficulties with the KG equation. 
In other words, the work adheres to the variational principle and to
results derived from it. Corrections that rely on other physical 
arguments are beyond the scope of this work. The variational
principle is applied 
here to quantum mechanics and to its classical limit as
well. Aspects of quantum field theory are pointed out at the end of
this work. Relativistic classical theory, relativistic quantum
mechanics and quantum field theory are related in an ascending
hierarchical order. A discussion of this kind of relationship between
physical theories can be found in [8]. The results of this work
provide new arguments supporting Dirac's point of view on the KG
equation. 

The second section discusses general properties of the KG equation. 
Problems belonging to the classical limit are discussed in the third
section. Concluding remarks are set out in the last section.

\vskip12pt
{\noindent \sectionfont 2. The Realm of Quantum Mechanics}
\vskip6pt

The Lagrangian density of the KG equation
of a free particle is (see [7], p. 26) 
\begin{equation}
{\mathcal L} = \frac {1}{2} (\phi_{,\mu }\phi _{,\nu} g^{\mu \nu} -
m^2 \phi^2).
\label{eq:KGFREE}
\end{equation}
Unless otherwise stated and
for the simplicity of the notation, the real KG field is examined
(see [7], p. 26). 
Applying the Euler-Lagrange equation (see, e.g. [7], p. 17)
\begin{equation}
\frac {\partial }{\partial x^\mu} \frac {\partial \mathcal L}
{\partial \phi _{,\mu}} - \frac {\partial \mathcal L}{\partial \phi} = 0.
\label{eq:EULAG}
\end{equation}
to the Lagrangian density  $(\!\!~\ref{eq:KGFREE})$, 
one obtains the KG equation of a free
particle $(\!\!~\ref{eq:KGEQ})$.

   Before entering into details, let us state the restrictions imposed by
the variational principle. The action $S$ is a Lorentz scalar and, in
units where $\hbar = 1$, it is a pure number. Thus, the
relation
\begin{equation}
dS = {\mathcal L} d^3xdt
\label{eq:ILDT}
\end{equation}
is examined. Since $d^3xdt$ is a Lorentz scalar having the dimensions $[L^4]$,
one finds that every term of the Lagrangian density should satisfy
the following requirements:
\begin{itemize}
\item[{A.}] It is a Lorentz scalar.
\item[{B.}] Its dimensions are $L^{-4}$.
\end{itemize}

First, it is proved that normalization problems of the KG equation still 
persist. Indeed, in examining $(\!\!~\ref{eq:KGFREE})$ and requirement B,
one finds that the dimension of the KG field $\phi $ is $[L^{-1}]$.
Thus, $\phi ^2$ has the dimension $[L^{-2}]$, which means that 
the corresponding wave function {\em cannot} represent probability
density whose dimension is $[L^{-3}]$. Therefore, the KG field 
$\phi $ lacks a fundamental property of a probability function.
This result holds for the complex KG field as well.

Another argument leading to this result is that $\phi $ is a
Lorentz scalar and so is $\phi ^*\phi$. On the other hand, the probability
density is a 0-component of a 4-vector called the 4-current
(see [9], pp. 69-71).
Hence, the Lorentz scalar $\phi ^*\phi$ cannot represent probability
density. This discussion provides new arguments supporting the well
known claim that the wave function of the KG equation cannot 
represent probability, which relies on the freedom of defining
$\partial \phi/\partial t$ (see [2], pp. 7,8).
Hence, the KG wave function
cannot be incorporated in the standard formulation of quantum mechanics,
where $\psi ^*\psi$ represents probability density and 
$\int \psi ^* {\hat O}\psi d^3x$
is the expectation value of any operator ${\hat O}$.

Next, it is proved that a KG particle cannot interact with an external 
electromagnetic field. Here real and complex KG fields are
treated separately. The analysis examines candidates for the interaction 
term of the Lagrangian density. Each of these 
candidates should satisfy 
requirements A and B. Moreover, due to
space homogeneity, quantities should not depend explicitly on the spatial
coordinates and, due to Lorentz covariance, they should not depend
explicitly on the time too. The electromagnetic equations of motion impose
further restrictions. Thus, varying the charged particle's coordinates
and holding the electromagnetic variables fixed, one should obtain 
the Lorentz 4-force $F^{\mu \nu}j_\nu $ exerted on a classical charge. Hence,
the interaction term of the Lagrangian density should be linear and
homogeneous in electromagnetic quantities. By the same token, 
applying a variation
of the electromagnetic field quantities, one should obtain Maxwell
equation $F^{\mu \nu}_{\;\;,\nu} = -4\pi j^\nu $. Therefore, the interaction
part of the Lagrangian density should also be proportional to a quantity
representing the 4-current of the investigated charged particle.
The electromagnetic fields tensor 
$F^{\mu \nu}$ is antisymmetric and is
unsuitable for this purpose. Indeed, due to requirement A,
it must be contracted with a second rank tensor depending on the KG field.
Two candidates are $\phi _{,\mu} \phi _{,\nu}$ and $\phi _{,\mu ,\nu}$.
However, these tensors are symmetric with respect to the indices
$\mu$ and $\nu $. Hence, their contraction with the antisymmetric
tensor $F^{\mu \nu}$ yields a null result.

Thus, the 4-potential of the electromagnetic fields, $A_\mu$, is left as
the sole electromagnetic factor of the required interaction term. In order
to have a Lorentz scalar it must be contracted with a 4-vector depending
on the KG field, which is $\phi _{,\mu }$. The dimension of $A^\mu $ is
$[L^{-1}]$ and that of $\phi _{,\mu }$ is $[L^{-2}]$. Hence, in
order to satisfy the dimensions of
requirement B, another factor $\phi$ should be added. Thus, the
candidate for the interaction term is 
\begin{equation}
{\mathcal L} _{int}= eA^\mu \phi_{,\mu} \phi
\label{eq:LINT}
\end{equation}
where the coefficient $e$ represent the dimensionless
elementary electric charge ($e^2 \simeq 1/137$).

Now applying the Euler-Lagrange equation $(\!\!~\ref{eq:EULAG})$
to the candidate for the interaction term
$(\!\!~\ref{eq:LINT})$, one obtains 3 terms
\begin{equation}
e(A^\mu \phi_{,\mu} + A^\mu _{,\mu} \phi - A^\mu \phi_{,\mu}).
\label{eq:EULINT}
\end{equation}
Here, the first and the last terms cancel each other and the second
term vanishes, due to the Lorentz gauge $A^\mu_{,\mu}=0$. 
This outcome means that an
external electromagnetic field does not affect the motion of a KG
particle defined by a real field. 

In the case of a complex KG field, one can write a conserved 4-current
(see [7], p. 40,[10], p. 30)
\begin{equation}
j_\mu = i(\phi ^*\phi _{,\mu} - \phi ^*_{,\mu}\phi ) 
\label{eq:JMU}
\end{equation}
and the following quantity 
\begin{equation}
{\mathcal L}_{int} = -ej^\mu A_\mu .
\label{eq:LCOMP}
\end{equation}
is tested as a candidate for the interaction part of the Lagrangian density.
(In $(\!\!~\ref{eq:JMU})$, the 4-current $j^\mu $ pertains to probability.
In other cases it represents an electric 4-current.) 
As required, the dimensions of $(\!\!~\ref{eq:JMU})$ and of
$(\!\!~\ref{eq:LCOMP})$ are $[L^{-3}]$ and $[L^{-4}]$, respectively.
Moreover, the interaction term $(\!\!~\ref{eq:LCOMP})$ is
proportional to the electric charge, 
which is a mandatory property of electrodynamics (see [9], p. 70).

Since $(\!\!~\ref{eq:JMU})$ is a conserved current, namely
$j^\mu_{,\mu} = 0$, one expects that
Maxwell equations hold for the interaction term $(\!\!~\ref{eq:LCOMP})$
(see [9], pp.  73-74). On the other hand, it is proved below that
problems exist with the KG equation.

   Applying the Euler-Lagrange equation $(\!\!~\ref{eq:EULAG})$ to
$\phi ^*$ and $\phi ^*_{,\mu }$
of the interaction term of the Lagrangian density
$(\!\!~\ref{eq:LCOMP})$, one obtains the interaction term of the
KG equation for $\phi $. Thus, the KG equation becomes
\begin{equation}
(\Box + 2ieA^\mu\partial _\mu + m^2)\phi = 0
\label{eq:KGEQEM}
\end{equation}

 Consider a motionless KG charged particle located inside a 
uniformly charged spherical shell. Thus, the 4-potential is
\begin{equation}
A_\mu = (V,0,0,0).
\label{eq:FOURP}
\end{equation}
Now, within the realm of quantum mechanics, the 
(unnormalized) wave function of a motionless KG particle is 
\begin{equation}
\phi = e^{-iEt}
\label{eq:ET}
\end{equation}
where the total energy is
\begin{equation}
E = m + U,
\label{eq:EMU}
\end{equation}
and the electrostatic energy is $U=eV$. In $(\!\!~\ref{eq:ET})$, the
omission of the spatial coordinates is an approximation. It is justified 
if the particle is enclosed in a sufficiently large spherical shell, where
the spatial derivatives yield negligible quantities.

  Thus, substituting $(\!\!~\ref{eq:FOURP})$-$(\!\!~\ref{eq:EMU})$, 
into $(\!\!~\ref{eq:KGEQEM})$, one finds 
\begin{equation}
[-(m + U)^2 + 2U(m + U) + m^2]\phi = U^2\phi \neq 0.
\label{eq:KGBAD}
\end{equation}
This result shows that the electrostatic interaction of the complex
KG field $(\!\!~\ref{eq:LCOMP})$ 
leads to a contradiction.

   An attempt to overcome this difficulty can be found in the
literature. For this purpose the interaction Lagrangian is 
(see [11], p. 275, [12], section 3)
\begin{equation}
{\mathcal L}_{int} = ie(\phi ^*_{,\mu} \phi - \phi^*\phi_{,\mu})A^\mu
- e^2A_\mu A^\mu \phi^* \phi.
\label{eq:SCHLI}
\end{equation}
Here a second term is added to the previously analyzed
interaction $(\!\!~\ref{eq:LCOMP})$.
However, unlike standard electromagnetic interactions of a charge $e$
with an external potential $A_\mu $, $(\!\!~\ref{eq:SCHLI})$
contains 2 terms: one is proportional to the electric charge
$e$ and the other is proportional to $e^2$

   One can see explicitly the problem emerging from this point, 
if the free electromagnetic term of the Lagrangian density 
\begin{equation}
{\mathcal L}_{EM-free} = - \frac {1}{16\pi} F^{\mu \nu }F_{\mu \nu }
\label{eq:LEMFREE}
\end{equation}
is added to the KG Lagrangian density
$(\!\!~\ref{eq:SCHLI})$. Varying the electromagnetic
potentials and their derivatives, one follows the standard treatment
(see [9], section 30) and finds the 
electromagnetic fields' equation associated with
a KG charge. Here, since $(\!\!~\ref{eq:SCHLI})$ contains 
an additional term
which is proportional to $e^2A^\mu A_\mu $, the fields' equation
associated with a KG charge is
\begin{equation}
F^{\mu \nu }_{\;\;,\nu } = -4\pi j^\mu + 8\pi e^2A^\mu  \phi^*\phi ,
\label{eq:KGMAXWELL}
\end{equation}
where $j^\mu $ is $e$ times the quantity
defined in $(\!\!~\ref{eq:JMU})$.

   Eq. $(\!\!~\ref{eq:KGMAXWELL})$ is inconsistent with Maxwell
equation
\begin{equation}
F^{\mu \nu }_{\;\;,\nu } = -4\pi j^\mu .
\label{eq:MAXWELL}
\end{equation}
Indeed, unlike Maxwell equation
$(\!\!~\ref{eq:MAXWELL})$, eq. $(\!\!~\ref{eq:KGMAXWELL})$
depends explicitly on the potentials. This property 
means that it is not gauge
invariant. Moreover, unlike Maxwellian fields whose inhomogeneous term
is proportional to the electric charge $e$, 
eq. $(\!\!~\ref{eq:KGMAXWELL})$ 
contains another term which is proportional to $e^2$.

The foregoing discussion completes the
proof that the KG field {\em cannot} interact with an external
electromagnetic field. This result means that a KG
particle cannot carry an electric charge.

   It is interesting to note that the Dirac field is free of the 2
discrepancies derived above for the KG field. Indeed, the Lagrangian
density of a free Dirac particle is (see [7], p. 54, [4], p. 102
and [6], p. 126)
\begin{equation}
{\mathcal L}_{Dirac} = \bar {\psi} (i\gamma ^\mu \partial _\mu - m)\psi.
\label{eq:LDD}
\end{equation}
Here the dimension of $\psi $ is $[L^{-3/2}]$ and that of 
$\bar {\psi} \psi $ is $[L^{-3}]$, as required for a function
representing probability density.
The electromagnetic interaction term of the Dirac field is the well known
quantity(see [7], p. 84, [4], p. 102
and [6], p. 135)
\begin{equation}
{\mathcal L}_{int} = \bar {\psi} (-e\gamma ^\mu A_\mu )\psi .
\label{eq:LDDINT}
\end{equation}
Indeed, $\bar {\psi }e\gamma ^\mu \psi $ is the Dirac
conserved 4-current (see
[10], p. 46, [12], pp. 23-24) and it is independent of electromagnetic
field quantities. Hence, $(\!\!~\ref{eq:LDDINT})$ is linear in $A_\mu $,
as required (see [9], section 30).

\vskip12pt
{\noindent \sectionfont 3. The Classical Limit}
\vskip6pt

   In the rest of this work it is shown that further problems arise
if one regards the KG field as a classical field (see [4]-[6]).
The first problem is examined for the simple case of a free KG particle.
Here the (yet unnormalized) wave function is assumed to take the form
\begin{equation}
\phi = e^{i({\bf p\cdot x} - Et)}.
\label{eq:PHIFREE}
\end{equation}
The action of a free classical particle and that of a free KG
particle are compared. It is proved that problems exist even for this
simple noninteracting solution of the KG equation.
Since for the KG particle we have a Lagrangian
density, there is a need for a definite expression for the
probability density. Here
the 0-component of the 4-current $(\!\!~\ref{eq:JMU})$ is used as the 
probability density $\rho $. Thus, for the free wave
$(\!\!~\ref{eq:PHIFREE})$, one finds that
\begin{equation}
\rho = 2E\phi ^*\phi
\label{eq:RHOFREE}
\end{equation}
From now on, it is assumed that the normalization of $\phi $ of
$(\!\!~\ref{eq:PHIFREE})$ satisfies
\begin{equation}
\int 2E\phi ^* \phi d^3 x = 1.
\label{eq:NORM}
\end{equation}

   Let us examine the action of an ordinary classical particle.
Here, one may use the Lagrangian (see [9], p. 25)
\begin{equation}
L = -m(1 - v^2)^{1/2}.
\label{eq:CLAG}
\end{equation}
Thus, the particle's action is
\begin{equation}
dS = -m(1 - v^2)^{1/2}dt.
\label{eq:CACTION}
\end{equation}
On the other hand, the Lagrangian density of a complex KG field is
(see [7], p. 38)
\begin{equation}
{\mathcal L} = \phi^*_{,\mu }\phi _{,\nu} g^{\mu \nu} -
m^2 \phi^* \phi.
\label{eq:KGFREECOMPLEX}
\end{equation}
Substituting $(\!\!~\ref{eq:PHIFREE})$ into $(\!\!~\ref{eq:KGFREECOMPLEX})$
and using the probability density $(\!\!~\ref{eq:RHOFREE})$,
one finds for the KG action
\begin{equation}
dS = [\int \frac {E^2 - p^2 - m^2 }{2E} \,(2E\phi ^* \phi)\, d^3x]dt= 0.
\label{eq:KGACTION}
\end{equation}
Hence, the action of the classical complex KG field 
$(\!\!~\ref{eq:KGACTION})$ is inconsistent with that of the
standard classical action $(\!\!~\ref{eq:CACTION})$.

Another contradiction arises if one examines the real
KG field and Einstein's equations
for the gravitational field (see [9], p. 276)
\begin{equation}
R^{\mu \nu} - \frac {1}{2}Rg^{\mu \nu } = 8\pi kT^{\mu \nu }.
\label{eq:GREQ}
\end{equation}
Here $T^{\mu \nu }$ is the energy-momentum tensor of matter and of the
electromagnetic fields and $k$ is the gravitational constant.

   The energy-momentum tensor can be derived from the Lagrangian
density in more than one way (see [9], pp. 77-80, 270-273). 
Thus, for the real KG field $\phi $,
the components of the energy-momentum tensor are
\begin{equation}
T^{\mu \nu} = \frac {\partial {\mathcal L}}{\partial \phi_{,\mu}}
\phi^{,\nu } - {\mathcal L}g^{\mu \nu},
\label{eq:EMT1}
\end{equation} 
Substituting the Lagrangian density $(\!\!~\ref{eq:KGFREE})$ into this
expression, one finds
\begin{equation}
T_{\mu \nu} = \phi _{,\mu } \phi _{,\nu } - 
\frac {1}{2}[(\phi_{,\alpha }\phi _{,\beta} g^{\alpha \beta} - 
m^2 \phi^2)]g_{\mu \nu }  
\label{eq:EMTKG}
\end{equation} 
(See [12], eq. (6) for the complex KG analog of this expression.)

Here the Yukawa field is examined (see [5], p. 21,[13])
\begin{equation}
\phi = g \frac{e^{-mr}}{4\pi r}.
\label{eq:YUKAWA}
\end{equation}
This expression is real and time independent. Now, the energy density is
the $T^{00}$ component of $(\!\!~\ref{eq:EMTKG})$. Hence, one finds for
the Yukawa field  $(\!\!~\ref{eq:YUKAWA})$
\begin{eqnarray}
T^{00}  & = & \frac {1}{2}[(\frac {g}{4\pi }\frac 
{\partial (e^{-mr}/r)}{\partial r})^2 + 
m^2\phi ^2] \nonumber \\ 
        & = & (\frac {1}{2r^2} + \frac {m}{r} + m^2)\phi ^2.
\label{eq:T00}
\end{eqnarray}

This result proves that the mass of a real KG particle does not behave like 
an ordinary mass. Indeed, the energy momentum tensor of an ordinary massive
particle is (see [9], p. 82)
\begin{equation}
T^{\mu \nu } = \frac {\mu }{\gamma } v^\mu v^\nu ,
\label{eq:TMASS}
\end{equation}
where the coefficient $\mu $ denotes mass density 
(to be distinguished from the index $\mu$) and 
$\gamma =(1 - v^2)^{-1/2}$ is the relativistic factor. Thus, the energy
momentum tensor of ordinary matter is proportional to the mass whereas
in the KG cases of $(\!\!~\ref{eq:EMTKG})$ and $(\!\!~\ref{eq:T00})$, 
it is a quadratic function of the mass.
This result proves that the mass of a real KG field is not an ordinary
mass.

Another point belonging to classical physics is the 4-force exerted on
a particle and the associated 4-acceleration. These notions are valid in
the validity domain of the classical limit and the 4-force is parallel
to the 4-acceleration. In the case of the
Yukawa field, the potential is the Lorentz scalar KG field
$(\!\!~\ref{eq:YUKAWA})$. Hence, the 4-force, which is a 4-vector,
is proportional to $\phi _{,\mu }$. Now, in the rest
frame of the source, the Yukawa field is time
independent and, in spherical coordinates, it depends only on
the radial coordinate $r$. Hence, $\phi $ of $(\!\!~\ref{eq:YUKAWA})$
yields $\partial \phi /\partial t = \partial \phi /\partial \theta =
\partial \phi /\partial \varphi = 0$.
These results prove that, at a given field point ${\bf r}$, the Yukawa
4-force takes the form
\begin{equation}
f^\mu = (0,\lambda \bf {r}),
\label{eq:4FORCE}
\end{equation}
where $\lambda $ is an appropriate coefficient.

On the other hand, the relation
\begin{equation}
v^\mu v_\mu = 1 \rightarrow v^\mu a_\mu = 0
\label{eq:VMUAMU}
\end{equation}
means that, in Minkowski space, the 4-acceleration (and the 4-force) must
be orthogonal to the 4-velocity. The Yukawa 4-force does not satisfy
this requirement. For example, take a particle moving towards the origin
of the Yukawa potential $(\!\!~\ref{eq:YUKAWA})$. Hence, the 4-velocity
of the particle takes the form
\begin{equation}
v^\mu = \gamma (1,-v{\bf r}/r).
\label{eq:4VELOCITY}
\end{equation}
Evidently, the scalar product of $(\!\!~\ref{eq:4FORCE})$ and 
$(\!\!~\ref{eq:4VELOCITY})$ does not vanish. This result proves that the
classical limit of the Yukawa potential is inconsistent with special 
relativity.

By contrast, in the case of electrodynamics, the Lorentz force density
is 
\begin{equation}
f^\mu = F^{\mu \nu }j_\nu ,
\label{eq:LORENTZFORCE}
\end{equation}
where $F^{\mu \nu }$ is the antisymmetric tensor of the fields. Since
the 4-current $j_\nu $ is parallel to the 4-velocity (see [9], pp. 70), one
realizes that the orthogonality requirement is satisfied.

\vskip12pt
{\noindent \sectionfont 4. Concluding Remarks }
\vskip6pt

   Several conclusions can be derived from the results obtained above.
The following remarks probably do not exhaust this issue.

   In electrodynamics there are two kinds of entities: massive
particles carrying charge and massless electromagnetic fields that
mediate interaction between charged particles. The KG particle may
be regarded as an entity that plays two roles. However, as
explained here, these roles are divided between the complex
and the real KG fields. Thus, the free complex KG field has a
conserved 4-current $(\!\!~\ref{eq:JMU})$ whereas the real KG
field lacks this property. Hence, the real KG field cannot describe
a free massive
particle. On the other hand, the Yukawa interaction term of the
Lagrangian takes the form (see [6], p. 135)
\begin{equation}
{\mathcal L}_{int} = g\bar {\psi} \psi \phi .
\label{eq:PSIPSIPHI}
\end{equation}
Now, like any other Hamiltonian, the associated Hamiltonian of
$(\!\!~\ref{eq:PSIPSIPHI})$
is a Hermitian operator and $\phi $ of this expression
must be real. Hence, the complex
KG field cannot be used as an interaction mediator.

This discussion shows that the real and the complex KG fields pertain
to two distinct physical tasks. The problems 
and contradictions of these fields, which
are derived above, are relevant to these tasks.

   As explained, the results hold for quantum mechanics and for its
classical limit. However, quantum field theory is a covering theory
of relativistic quantum mechanics and of its nonrelativistic version[8]. 
It is well known that there is a close
connection between the Dirac field of quantum field theory and 
the Dirac's equation in relativistic quantum mechanics and
of its nonrelativistic approximation as well. 
This work shows that, in the case of a
KG particle, problems exists within the realm of 
classical and quantum mechanics.
Hence, in the case of a KG particle, it is necessary to
explain the transition from quantum field theory to
quantum mechanics.

   It is proved in Section 2 that the dimension of 
a KG field $\phi ^* \phi$
is $[L^{-2}]$ and that it is unsuitable for representing density of a
KG massive particle. The same conclusion is obtained from the fact
that $\phi ^*\phi$ is a Lorentz scalar.
The density $j_0$ of the conserved 4-current of
the complex KG field $(\!\!~\ref{eq:JMU})$ is a sum of terms 
that take the form $\phi ^*_{,\mu} \phi$. This form 
differs from the ordinary quantum mechanical 
expression of probability, which is
the square of the absolute value of a wave function.

   The $0^-$ $\pi $ mesons cannot be regarded as KG particles. Indeed,
charged $\pi $ mesons
are found in a free state, very far away from the interaction
region and quantum mechanics as well as its
classical limit hold in this case. 
Thus, since it is proved in Section 2 that a KG particle
cannot carry electric charge, one concludes that the $\pi ^\pm $
mesons are not KG particles. Isospin symmetry extends this conclusion to
the $\pi ^0$ meson.

This conclusion is supported by the following general argument. It is
now known that a $\pi $ meson is not an elementary structureless particle
but a composite system that contains a quark and an antiquark. Hence, a
field function $\phi (x^\mu)$ which depends on a 
{\em single} set of 4 coordinates
$x^\mu $ may be relevant to the center of mass coordinates of the
system but it {\em cannot} describe its internal degrees of freedom.

   Further problems arise if the KG field is regarded as a classical
field. Here, as shown in Section 3,
the action of a complex KG field, the mass parameter 
of the real KG Lagrangian density and the Yukawa force disagree with
the corresponding classical quantities.

   On the other hand, it should be stated that this work does not
deny the usage of the
KG equation as a {\em phenomenological equation}. Indeed,
by definition, a phenomenological equation is evaluated mainly (or only)
by its usefulness in describing a specific set of data. This kind
of evaluation is of a practical nature and is immune to theoretical
counter-arguments. The case of the $\pi $ mesons illustrates this
issue. Thus, in low energy experiments, where
excitations of the $0^-$ quark-antiquark 
ground state can be ignored, a $\pi $ meson may
be regarded as an elementary object and the KG equation may
be used phenomenologically. This approach pertains also to bound states 
of a proton and a $\pi ^-$ meson.

This work provides an example of the strength of the variational
principle. Thus, if one adheres to it then restrictions on the
validity of physical theories may arise. Here it is shown that 
if this approach is applied then the KG
equation encounters contradictions in the standard relations between
wave function and probability, in electromagnetic interactions and
in classical aspects of the problem.

Acknowledgment: I thank Prof. L. Horwitz for carefully reading the
manuscript and making useful remarks.

\newpage
References:
\begin{itemize}

\item[{[1]}] Feshbach H and Villars F 1958 Elementary Relativistic
Wave Mechanics of Spin 0 and Spin 1/2 Particles {\em Rev. Mod. Phys.}
{\bf 30}, 24-45
\item[{[2]}] Weinberg S 1995 The Quantum Theory of Fields
(Cambridge: University Press). Vol. 1. 
\item[{[3]}] Dirac P A M 1978 Mathematical Foundations of Quantum
Theory Ed Marlow A R (New York: Academic). (See pp. 3,4). 
\item[{[4]}] Barut A O 1965 Electrodynamics and Classical Theory
of Fields and Particles (New York: MacMillan). See pp. 102, 211.
\item[{[5]}] Mandl F 1965 Introduction to Quantum Field Theory
(New York: Interscience). P. 13.
\item[{[6]}] Sterman G 1993 An Introduction to Quantum Field Theory
(Cambridge: University Press). P. 7.
\item[{[7]}] Bjorken J D and Drell S D 1965 Relativistic Quantum
Fields (New York: McGraw).(See chapter 12.)
\item[{[8]}] Rohrlich F 1965 Classical Charged Particles
(Reading Mass: Addison-wesley). pp. 3-6. 
\item[{[9]}] Landau L D and Lifshitz E M 1975 The Classical
Theory of Fields (Oxford: Pergamon)
\item[{[10]}] Ryder L H 1985 Quantum Field Theory (Cambridge University
Press: Cambridge)
\item[{[11]}] Schweber S S 1964 An Introduction to Relativistic
Quantum Field Theory (New York: Harper).
\item[{[12]}] Pauli W and Weisskopf V 1934 The Quantization of the
Scalar Relativistic Wave Equation {\em Helv. Phys. Acta} {\bf 7} 709-731.
English translation: Miller A I 1994 Early Quantum
Electrodynamics (Cambridge: University Press). pp. 188-205.
\item[{[13]}] Bjorken J D and Drell S D 1964 Relativistic Quantum
Mechanics (New York: McGraw).(See p. 211.)
\end{itemize}

\end{document}